# Thermal conductivity of graphene kirigami: ultralow and strain robustness

Ning Wei,[1] Yang Chen,[1] Kun Cai,[1] Hui-Qiong Wang,[3,4] Junhua Zhao*[2] and Jin-Cheng Zheng*[3,4,5]

1 College of Water Resources and Architectural Engineering, Northwest A&F University, Yangling 712100, China

2 Jiangsu Key Laboratory of Advanced Food Manufacturing Equipment and Technology, Jiangnan University, Wuxi, China

3 Department of Physics, Xiamen University, Xiamen City, Fujian Province, China 361005

4 Xiamen University Malaysia Campus, 439000 Sepang, Selangor, Malaysia

5 Fujian Provincial Key Laboratory of Mathematical Modeling and High-Performance Scientific Computation, Xiamen University, Xiamen City, Fujian Province, China 361005

Kirigami structure, from the macro- to the nanoscale, exhibits distinct and tunable properties from original 2-dimensional sheet by tailoring. In present work, the extreme reduction of the thermal conductivity by tailoring sizes in graphene nanoribbon kirigami (GNR-k) is demonstrated using nonequilibrium molecular dynamics simulations. The results show that the thermal conductivity of GNR-k (around 5.1 Wm-1K-1) is about two orders of magnitude lower than that of the pristine graphene nanoribbon (GNR) (around 151.6 Wm-1K-1), while the minimum value is expected to be approaching zero in extreme case from our theoretical model. To explore the origin of the reduction of the thermal conductivity, the micro-heat flux on each atoms of GNR-k has been further studied. The results attribute the reduction of the thermal conductivity to three main sources as: the elongation of real heat flux path, the overestimation of real heat flux area and the phonon scattering at the vacancy of the edge. Moreover, the strain engineering effect on the thermal conductivity of GNR-k and a thermal robustness property has been investigated. Our results provide physical insights into the origins of the ultralow and robust thermal conductivity of GNR-k, which also suggests that the GNR-k can be used for nanaoscale heat management and thermoelectric application.

## Introduction

Graphene, the classic of two-dimensional material, has attracted intensive research interests recently due to its outstanding mechanical, thermal, and electronic properties.[1-7] The thermal conductivity of suspended monolayer graphene was measured to be as high as 3000~5300 W m$^{-1}$ K$^{-1}$.[5,8,9] The tensile stiffness and strength of graphene sheet are on the order of 1 TPa and 100 GPa,[6] respectively. Thus, graphene has been considered as one of the best candidate materials for the post-COMS (complementary metal-oxide-semiconductor)[10] and flexible electronic device technology[11,12] due to all its advanced properties. In addition, the properties of graphene can be adjusted on demand by doping,[13] strain,[14] defects,[15-18] chemical functionalization[19] and tailoring,[20] (such as, the thermal conductivity of graphene is decreased by ~50% with Stone-Wales defects,[17] ~60% with strain[14] and ~80% with chemical functionalization.[19]) etc, for a variety of purposes. For instance, one can maximize thermal conductance for cooling applications or minimize conduction in thermoelectric applications.

Recently, a new nanostructure of graphene, a typical patterned structure, graphene nanoribbons kirigami (GNR-k) has been obtained by tailoring graphene in experiment, which exhibits outstanding and tunable mechanical properties.[21] The properties of GNR-k are flexible and tunable with tailoring on demand. The tailored GNR-k structures are helpful to improve ductility and brittle behavior.[22,23] The GNR-k structures exhibit strong yield and fracture strains that can be up to three times higher than that of the pristine graphene nanoribbon (GNR) using molecular dynamics simulations (MD) according to the latest study.[22] Although the mechanical properties of GNR-k have been studied, their thermal conductivity of GNR-k and the effect of their geometry parameters are still not clear. The cutting strip holes of GNR-k suggest a low thermal conductivity due to the effect of nanoconstrictions and ballistic resistances.[24] Combining the tailoring effect on the mechanical and the thermal properties of GNR-k, the influence of the tensile strain on the thermal conductivity of GNR-k is quite different with that of GNR. The reason is that the kirigami structure turns the tensile strain to the geometry deformation and the carbon bonds of GNR-k are barely stretched, which is completely different with those of the GNR under tension.[22]

To apply the GNR-k into potential flexible micro/nano electronic devices, it is significant and necessary to understand the mechanism of its low thermal conductivity and the effect of the strain and geometry parameters on its conductivity.

In this work, we have investigated the thermal conductivity of GNR-k with various tailored nanostructures by nonequilibrium MD simulations. The ultralow thermal conductivity of GNR-k has been found and the minimum value is expected to be approaching zero in extreme case from our theoretical model. The mechanism of the reduction of the thermal conductivity has also been revealed by the elongation of real heat flux path, the overestimation of real heat flux area and phonon scattering at the vacancy of the edge. In particular, the effect of the tensile strain on the thermal conductivity of GNR-k indicates that a robustness of thermal conductivity, namely, independent of the given tensile strains, can be realized by controlling their geometry parameters. Finally, our MD results show that the ultralow and robust thermal conductivity of GNR-k makes it to be a potential thermoelectric or a thermal management material.

## Modeling & Method

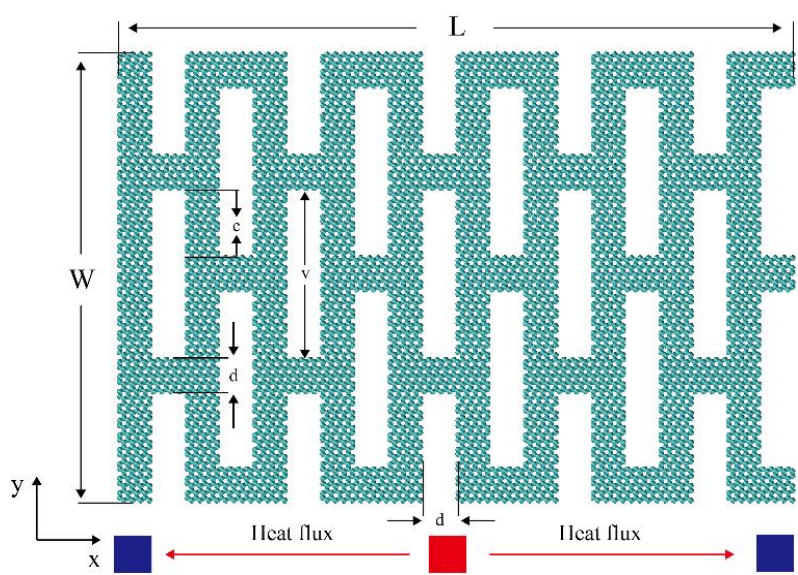

**Fig.1** The schematic and its geometry parameters of graphene nanoribbon kirigami.

The atomic structure and its geometry parameters of GNR-k were depicted in **Fig.1**. The width and the length of GNR-k were defined as *W* and *L*, respectively. The heat flux transports along the *x* direction (zigzag direction). To eliminate the size effect induced difference on the thermal conductivity, the length *L* of all the GNR-k samples is identical and is equal to 19.4 nm, while the width *W* is different with different tailoring sizes. The values of all geometry parameters were listed in **Table 1**. The distance between the interior cuts and the width of the tailored ribbons was defined as *d* and was fixed as 1 nm in all our MD simulations. The width of the tailoring vacancy is *v*. The wingspan *c* was defined as *c* = (*v* - *d*)/2.

**Table 1**. All geometry parameters for GNR and GNR-k.

| Parameters | GNR | GNR-k1 | GNR-k2 | GNR-k3 | GNR-k4 | GNR-k5 |
|---|---|---|---|---|---|---|
| *W* (nm) | 7 | 5 | 7 | 9 | 11 | 13 |
| *v* (nm) | 0 | 1 | 2 | 3 | 4 | 5 |
| *c* (nm) | / | 0 | 0.5 | 1 | 1.5 | 2 |
| No. atoms | 5440 | 2960 | 3760 | 4440 | 5240 | 5960 |

All MD simulations were performed to explore the mechanical and thermal transport properties of GNR-k using the large-scale atomic/molecular massively parallel simulator (LAMMPS).[25] Periodic boundary conditions were applied on the *x* direction (see **Fig.1**). The adaptive intermolecular reactive empirical bond-order (AIREBO) potential,[26] was employed to describe the interatomic interaction between carbon atoms. To avoid the spurious strengthening effect and the nonphysical part in the tensile fracture process, the cutoff parameter was set to be 0.2 nm for the REBO part, as suggested by reference.[27] The atomic structures of GNR-k were optimized using the Polak-Ribiered version of conjugated-gradient algorithm initially.[28] Then, 200 ps Nose-Hoover thermal bath (coupling constant 0.1 ps) were conducted to make the system reach equilibrium state at 300K (the time step is 0.5 fs).[29,30] To study the tensile strain effect on the thermal conductivity of GNR-k, the uniaxial engineering strain along the *x* direction was performed under the deformation-control method with strain rate of 0.001/ps in a micro-canonical ensemble and the strain increment was applied every 1000 time steps.[14] The engineering stress is defined as

$$\sigma_x = \frac{1}{V_0}\frac{\partial U}{\partial \varepsilon_x}, \quad (1)$$

where *U* is the strain energy, $V_0$ is the initial volume of the system. The thickness of graphene was taken as 0.34 nm in the mechanical calculations.[31] The atomic stress of individual carbon atoms in the graphene sheet is calculated according to the equation,

$$\sigma_{ij}^{\alpha} = \frac{1}{\Omega^{\alpha}}\left(\frac{1}{2}m^{\alpha}v_i^{\alpha}v_j^{\alpha} + \sum_{\beta=1,n}r_{\alpha\beta}^{j}f_{\alpha\beta}^{i}\right), \quad (2)$$

where *α* and *β* are the atomic indices; *i* and *j* denote indices in the Cartesian coordinate system, $m^{\alpha}$ and $v^{\alpha}$ denote the mass and velocity of atom *α*, $r_{\alpha\beta}$ is the distance between atom *α* and *β*. The second term sums over all atoms and incorporates the contributions of kinetic energy, pairwise and many-body interactions. The stress on each atom was averaged over the last latter 500 time steps of the relaxation period. The global stress of the system was then obtained by averaging the stress on each atom over all.

The thermal conductivity was obtained by a reverse non-equilibrium MD method, in which the heat flux is imposed on the system to form a temperature gradient.[32] Along the heat transfer direction, the studied system was partitioned into 50 thin slabs of equal thickness, the heat source (the 26th slab) and sink slabs (the 1st slab) were located at the middle and the ends of the model (see **Fig.1**). Then, the heat flux $J$ was subtracted from or injected into these two slabs by exchanging the kinetic energies (every 100 time steps) between the coldest atom (with the lowest kinetic energy) in the heat sink slab and the hottest atom (with the highest kinetic energy) in the heat source slab. The heat flux $J$ can be calculated from the exchanging amount of the kinetic energy between two selected layers according to the following equations

$$J = \frac{\sum_{Nswap} \frac{1}{2}(mv_h^2 - mv_c^2)}{t_{swap}}, \tag{3}$$

the temperature distribution was collected after 1 ns (with time step 0.5 fs) and the non-equilibrium steady state was reached. The temperature of each slab is expressed as follows

$$T_i(\text{slab}) = \frac{2}{3Nk_B} \sum_j \frac{p_j^2}{2m}, \tag{4}$$

where $T_i$ (slab) is the temperature of the $i$th slab, $N$ is the number of carbon atoms in this slab, $k_B$ is Boltzmann's constant and $p_j$ is the momentum of atom $j$. The temperature profiles were obtained by averaging results of last 6 million time steps (3 ns) which are collected every 100 time steps. Two typical examples of temperature distribution of GNR and GNR-k (with $v$ = 5nm) were shown in **Fig.2** and the temperature gradient, d$T$/d$x$, was obtained by linear fitting in labeled region. The curve of the temperature profile for GNR-k exhibits more roughness than that of GNR. The probable reason is that GNR-k can not be treated as a homogenous material in such a small size and its localized vacancies affect the heat transport (as discussed in heat flux transport section) and temperature distribution.

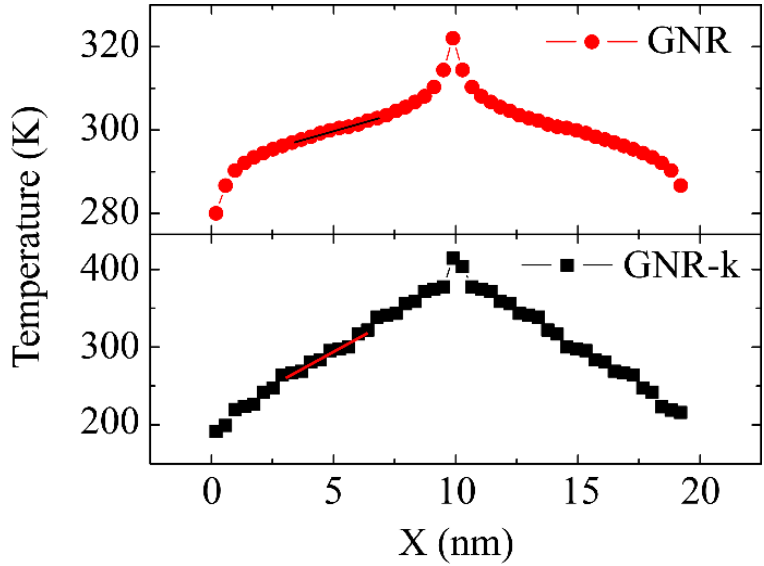

Fig. 2 The temperature profile of GNR (upper) and GNR-k (lower) (the linear fitting region is labeled with line segment).

The thermal conductivity is then calculated using the Fourier law:

$$\kappa = \frac{J}{2A\, \partial T/\partial x},\qquad(5)$$

where $A$ is the cross section of heat transfer, defined by the width times the thickness of a single C-C bond length (0.14 nm) of the GNR.[33]

## Results & Discussion

We first studied the effect of the vacancy size $v$ on the thermal conductivity of GNR-k at room temperature ($T$ = 300 K) (Note that the graphene samples being discussed in this study only refers to zigzag chiral, except where noted). The normalized $\kappa/\kappa_0$ with different $v$ from 0 (pristine GNR) to 5 nm using MD simulations were shown in **Fig. 3a**, while the distribution of $\kappa/\kappa_0$ with various $v$ by present theoretical model when $v >$5 nm is plotted in **Fig. 3b**, where $\kappa_0$ and $\kappa$ represent the thermal conductivity of GNR and the different tailored GNR-k in **Table 1**, respectively. The error bar is obtained from the linear fitting of the temperature gradient, d$T$/d$x$, which represents the standard error of $\lambda$ according to Eq. (3). The thermal conductivity of the perfect GNR in **Table 1** is around 151.6 W·m$^{-1}$k$^{-1}$, while the thermal conductivity of GNR-k5nm in Table 1 is only approximate to 5.1 Wm$^{-1}$k$^{-1}$ using present MD simulations. The maximum reduction can be up to 97% when $v$ is equal to 5 nm by comparison of the corresponding GNR in **Fig. 3a**. The thermal conductivity of GNR-k decreases slowly when $v$ is larger than 3 nm (a reduction is only around 2% from 3 nm to 5 nm), while it decreases sharply with increasing $v$ from 1 nm to 3 nm (a reduction is up to 12% from 1 nm to 3 nm). The chiral and hydrogen atoms termination effects were calculated for GNR-k2nm, as shown in **Fig. 1A** of Appendix. Since all of those simulations in **Fig. 1A** gave the same tendency of thermal conductivity and we mainly focus on the geometry factor and the strain effect on the thermal conductivity of GNR-k, the effect of the chirality and the hydrogen termination on the thermal conductivity was neglected in the following MD simulations.

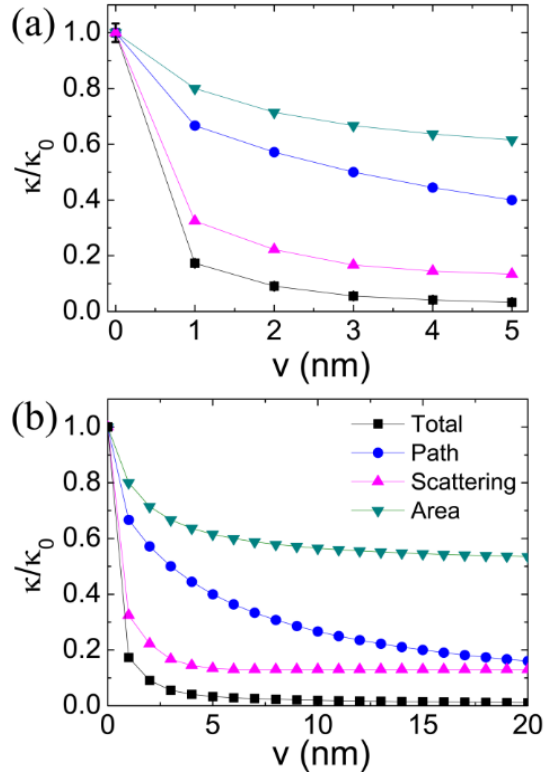

**Fig. 3** The relative thermal conductivity ($\kappa/\kappa_0$) of GNR-k as a function of the vacancy size $v$. The thermal conductivity of GNR is defined as $\kappa_0$, which is equal to 151.6 W·m$^{-1}$k$^{-1}$ here. The effect of the elongation of real heat flux path and the phonon scattering are analyzed independently. The MD results are shown in upper (a); the corresponding theoretical studies are exhibited in lower (b).

To clarify the mechanism for the reduction of the thermal conductivity of GNR-k, the spatial distribution of the heat flux by vector arrows on each atom in GNR-k5nm under non-equilibrium steady state was shown in **Fig.4**. The atomic heat flux is defined from the expression: $\mathbf{J}_i = e_i\mathbf{v}_i - \mathbf{S}_i\mathbf{v}_i$, where $e_i$, $\mathbf{v}_i$, and $\mathbf{S}_i$ are the energy, velocity vector and stress tensor of each atom $i$, respectively.[15] It can be obtained by calculating the atomic heat flux in the reverse non-equilibrium MD simulations and the results were averaged over 2 ns in the steady state (1 ns after the MD simulations). The vector arrows show the migration of the heat flux on surface of GNR-k and reflect vividly the elongation of heat flux path and the phonon scattering around the vacancy regions. The phonon scattering occurs at the vacancy regions of the tailored edges (see **Fig. 4**). When a propagating phonon tries to pass through a vacancy barrier in GNR-k, it will be scattered at the boundary and turn its direction to along the ribbons, which results in the reduction of the thermal conductivity. Moreover, the real heat flux path with red arrows of GNR-k (see **Fig. 4**) is greater than that of the perfect GNR, while the real area of GNR-k is lower than that of GNR. Both the reasons cause the reduction of thermal conductivity.

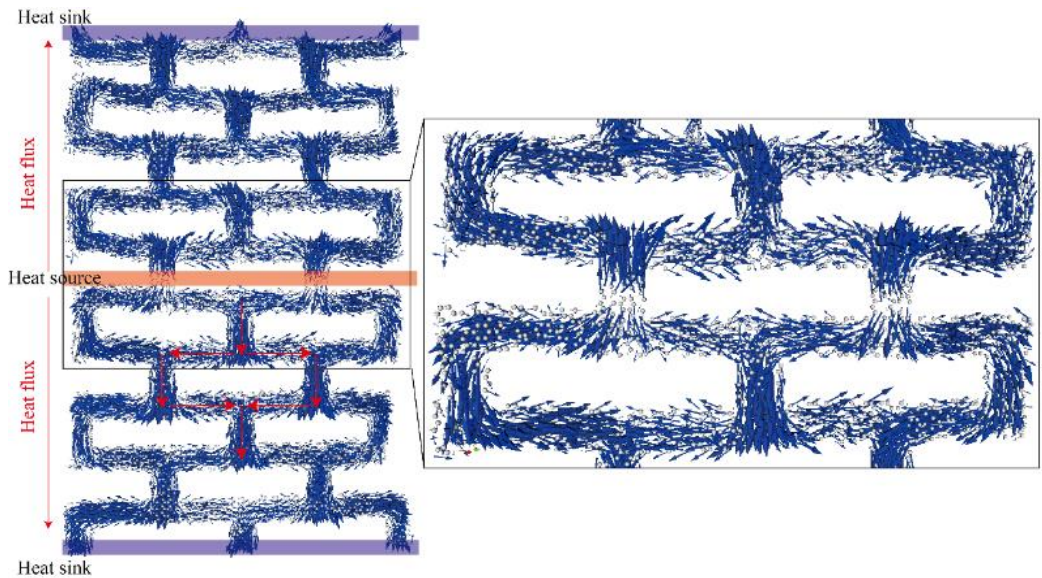

**Fig. 4** (color online) Spatial distribution of heat flux by vector arrows on each atom in GNR-k5 under non-equilibrium steady state. The micro heat fluxes are obtained under non-equilibrium steady state with the heat source and sink are located at the middle and both ends of the model. The global heat flux transport directions are labeled with red arrows at the left side of GNR-k and the real heat flux transport paths are labeled on the graphene ribbons. The zoom in viewport is plotted at the right side.

To understand the contribution of the three main reasons on the reduction of the thermal conductivity, the reduction from the phonon scattering, the effective area and the elongation of the heat path can be defined as the reduction parameters of $\delta_s$, $\delta_a$ and $\delta_p$, respectively. The thermal conductivity of GNR-k can be expressed as:

$$\kappa = \delta_s \cdot \delta_a \cdot \delta_p \cdot \kappa_0, \qquad (6)$$

where $\kappa_0$ is the thermal conductivity of pristine GNR. As shown in **Fig. 4**, the real heat flux path of

GNR-k is larger than that of the perfect GNR, while we use the same path length of GNR to calculate the d$T$/d$x$ of GNR-k. Therefore, the d$T$/d$x$ of GNR-k is overestimated, which results in the reduction of the thermal conductivity of GNR-k from Eq.(3). For the contribution of the elongation of real heat flux path, the reduction parameter $\delta$ can be defined from the geometry structure of GNR-k as: $\delta_p = 1/(1.25+ v/4b)$. On the other hand, the heat flux area $A$ (here $A=W \times h$, and $h$ is the thickness 0.14 nm) of GNR-k is lower than that of the perfect GNR, while we use the same heat flux width $W$ of GNR to calculate thermal conductivity of GNR-k in Eq.(3). Therefore, the heat flux width of GNR-k is also overestimated, which is another reason to cause the reduction of the thermal conductivity of GNR-k. The reduction parameter of area is $\delta_a = (v+3b)/(2v+3b)$ and it converges to 0.5 in **Fig. 3b** when $v$ tends to infinity. The thermal conductivity κ can be obtained from the MD simulations, and then the contribution of phonon scattering $\delta_s$ can be calculated from $\delta_s=\kappa/(\delta_a \cdot \delta_p \cdot \kappa_0)$. The two parameters, $\delta_a$ and $\delta_p$, are easily calculated from above definition. The rest phonon scattering parameter, $\delta_s$, is converged to 0.13 when $v$ is large enough ($v$ ≥5nm, see **Fig. 3b**), in which the heat flux transport on GNR-k is similar to on a narrow graphene ribbon. As shown in **Fig. 3b**, the thermal conductivity of GNR-k could be almost zero when $v$ is up to 20 nm. The effect of heat transport area $\delta_a$ and the phonon scattering $\delta_s$ decreases with increasing $v$ and converges to a constant value, respectively, while $\delta_p$ always keeps decreasing. The parameter $\delta_p$ is larger than $\delta_s$ when $v$ is higher than 26 nm. In particular, the effect of the elongation of real heat flux path is the main reason for the reduction of thermal conductivity for the large system, such as in micro/nano scale in experiment. In addition, we performed a supplemental simulation with $v$ = 5nm and the vacancy parallel to the thermal transport direction, as shown in **Fig. 2A** of Appendix. Our MD simulation shows that the value of the thermal conductivity is 40.42 Wm$^{-1}$K$^{-1}$, which is about 8 times higher than that of the perpendicular vacancy model. Both parallel and perpendicular vacancy models share the same vacancy area but have different thermal transport properties. The parameter $\delta_s$ of parallel model is 0.55 which is close to the scattering parameter of perpendicular model of $v$ = 1nm ($\delta_s$ = 0.34). The value of path elongation reduction parameter $\delta_p$ is 0.80 which is twice as large as the perpendicular one. The parallel results indicate that the thermal conductivity of GNR-k display anisotropic properties.

From above analysis, three main reasons which result in the reduction of the thermal conductivity could be summarized as follows: (1) The real heat flux path of GNR-k is underestimated; (2) The real heat flux area (or width) of GNR-k is overestimated; (3) The phonon scattering is occurred at the tailored edges of vacancy regions.

To study the potential application of GNR-k as the flexible electronic devices, we further studied its mechanical properties and the strain effect on its thermal conductivity at $T$=300 K. **Fig.5** (see **a1-a5** and **b1-b5**) shows the stress distribution on each atom of GNR-k under different tensile strain along the zigzag direction when $v$ = 2 nm and 5 nm, respectively. The snapshots were scaled to the same length for the illustration. **Fig.5** (see **c1** and **c2**) shows the side view of the stretched GNR-k at tensile strain of 20% and 80% for $v$ = 2 nm and 5 nm, respectively, in which the two strains are both close to their fracture strains. The oblique view of GNR-k2nm is shown in **Fig.5d**. The out-of-plane rotation of part $c$ in GNR-k was occurred during the loading. Same phenomenon could be observed in the loading of real kirigami paper.[22] **Fig.5 (e1)** shows the stress-strain relationship of GNR-k with different $v$ from 0 nm (perfect GNR) to 5 nm. The corresponding fracture stresses and fracture strains were shown in **Fig.5 (e2)**. The fracture stresses decrease with increasing $v$, while the fracture strains with different $v$ exhibit the opposite tendency (without considering the perfect graphene). The tailored GNR-k mechanical behavior is more like a vacancy defect graphene when $v$ is 1 nm, in which the fracture strain reduces to ~10%. When $v$ is higher than 3 nm, the extensibility of fracture strain has a great advance which increases from 20% (prefect graphene) to 40%. The main reason is that the behavior of GNR-k is close to a vacancy defective graphene for a small $v$. On the other hand, for large $v$, its loading strain is released by the geometry deformation (see **Fig.5 (b1-b5)**). There are two stages for high $v$ ($v \geq 3$ nm) in the tensile process. In the first stage, the GNR-k can readily elongate up without some remarkable stresses because of the GNR-k geometry deformation. In other words, the GNR-k is extremely flexible with nearly zero stiffness (as shown in **Fig. 6a**). In the second stage, the tensile stress increases sharply with increasing tensile strain until the final fracture is occurred. When the tensile strain is large enough, the parts of $c$ (which is perpendicular to the tensile direction in **Fig.1**) converts into parallel to the tensile direction (see **Figs.5 (a5)** and **(b5)**). From the stress distribution, we can find that the elongation on each carbon-carbon bond is not uniform and only the inner side edges bonds are sharply deformed. The radial distribution function (RDF) for GNR and GNR-k with various strains was shown in **Fig. A3** of Appendix. The bond length is almost uniform with different tensile strains for both $v$ = 2 nm and 5 nm samples of GNR-k, respectively, in which most of bonds are barely elongated and the average bond length slightly changes under different tensile strains.

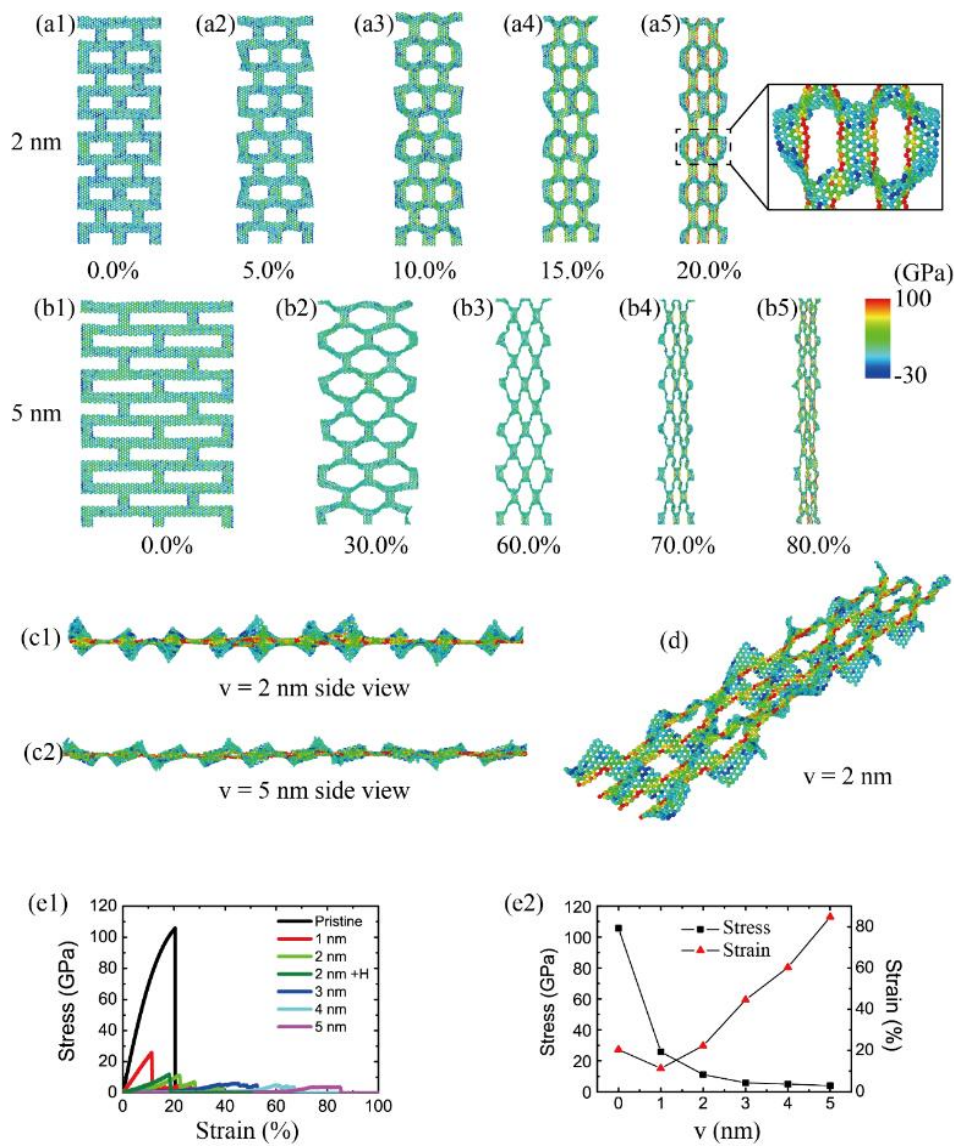

**Fig. 5** The configurations and mechanical properties of GNR-k under tensile strains. (a1-a5) The snapshots of the deformed GNR-k2nm under various tensile strains (labeled under the corresponding configurations); (b1 - b5) The snapshots of the deformed GNR-k5nm (all the snapshots were scaled to the same length for the illustration); (c1) the side view of the stretched GNR-k at tensile strain of 20% (close to the fracture strain) for $v$ = 2 nm; (c2) the side view of the stretched GNR-k at tensile strain of 80% (close to the fracture strain) for $v$ = 5 nm; (d) An oblique view of GNR-k2nm; (e1, e2) The mechanical properties of GNR-k with various tailoring size $v$.

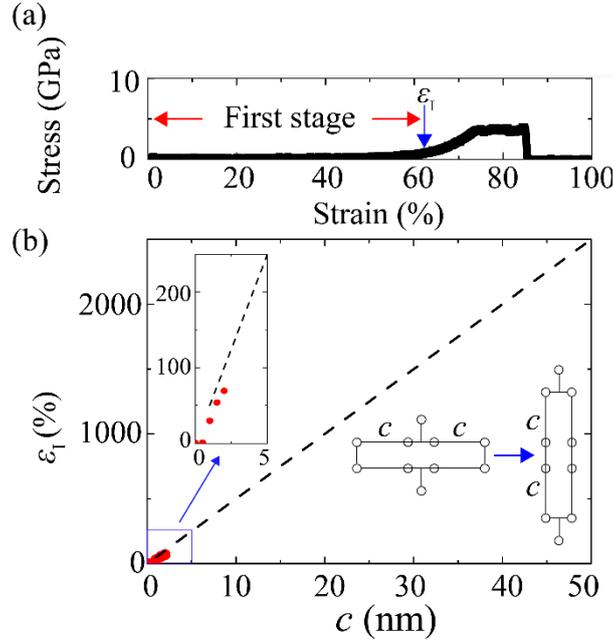

**Fig. 6** (a) The stress-strain curves of $v$ = 5 nm and the definition of the first state of the loading and its maximum strain $\varepsilon_I$; (b) The relationship between the maximum strain $\varepsilon_I$ and $c$ based on present theoretical model, in which $c$ is defined as the width of wingspan $c = (v - b)/2$. (The dash line is plotted with equation: $\varepsilon_I = c/2b \times 100\%$ and the red dots are from the MD results.)

To analyze the mechanism of the effect of $v$ on the mechanical behavior of GNR-k, the maximum strain of the first stage was defined as $\varepsilon_I$ (see **Fig. 6a**). **Fig. 6a** shows that $\varepsilon_I$ increases with increasing $v$. When the tensile strain is lower than $\varepsilon_I$, the tensile strain mainly induced the geometry deformation, as shown in **Fig.5**. To obtain the relationship between $\varepsilon_I$ and $v$, we simplified the kirigami structure to a simple point-stick model (see **Fig. 6b** inset). It should be noted that the energy of the bond angle is ignored here. Wingspan of kirigami, $c$, (see **Fig.1**) is the free part can be easily convert into parallel to the tensile direction at the end of the first stage. To quantitatively describe the relationship between $\varepsilon_I$ and $c$, we can define $\varepsilon_I = c/2b \times 100\%$ under the ideal condition for a theoretical model. When $c$ < 1.0 nm, $v$ is less than 2.0 nm in present model and the behavior of GNR-k is close to the vacancy defects, where $\varepsilon_I$ is setting to 0.0. When $c$ is greater than 1.0, the MD simulation results are in good agreement with the theoretical prediction in **Fig.6b**. The present results show that a high ductility GNR-k can be designed according to the tailoring patterns.

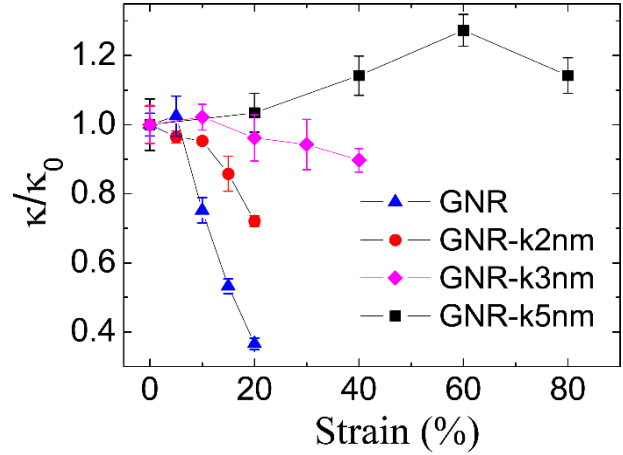

**Fig. 7** Thermal conductivity (reduced with thermal conductivity value of stress free as 1.0) as a function of strain.

In this section, the strain effect on the thermal conductivity of GNR-k was studied using present MD simulations in **Fig. 7**. For the case of GNR, its thermal conductivity has a slight variation with small strain (<5%) and then decreases remarkably with increasing strain (>5%). The reduction of the thermal conductivity of GNR under tensile strain should be attributed to the softening of phonon modes and the increase of lattice anharmonicity.[14] A maximum of 65% reduction is achieved before fracture failure occurred. This trend is in good agreement with previous studies.[34] For the cases of GNR-k, the relationship between the thermal conductivity and the tensile strain nonlinearly changes with the tailoring size $v$. The thermal conductivity of GNR-k2nm is still sensitive to the tensile strain and it decreases rapidly (30% reduction at 20% tensile strain), while the thermal conductivity is insensitive to the tensile strain when $v \geq$ 3nm ($c >$ 1.0). The value of the thermal conductivity decreases less than 10% for $v$ = 3nm until the system is destroyed. Furthermore, the thermal conductivity increases in the whole tension process for $v$ = 5nm. The two possible reasons results in the tendency of the thermal conductivity and are summarized as follows: First, the tensile deformation causes the reduction of the thermal conductivity of perfect GNR for softening bonds and induces the interphonon scattering.[14] On the other hand, the tensile deformation increases the length of the system, which should cause the increase of the thermal conductivity. The initial length is always equal to 19.4 nm in our MD simulations which is much shorter than the phonon mean free path of graphene (~1μm).[35] The thermal conductivity of GNR increases with increasing system length according to a power law (the divergent exponent is 0.35 for zigzag graphene).[36] Therefore, the distribution of the thermal conductivity with different tensile strain was determined by the competition of the two mechanisms. When $v$ is higher than 3 nm ($c >$ 1.0), one can find the first stage (the strain is smaller than $\varepsilon_I$ in **Fig.6a**) in the tensile strain without inducing extra-stress/force of GNR-k. Therefore, the structures of GNR-k display a great ductility. The values of $\varepsilon_I$ for GNR-k3nm and GNR-k5nm can be up to 20% and 60%, respectively (see **Fig.6a**). The thermal conductivity slightly increases for GNR-k3nm when the strain is less than 20%, while it slowly decreases around 10% with increasing strain (>20%) before the failure. For the case of GNR-k5nm, the thermal conductivity slowly increases with increasing strain (<60%) and then decreases (>60%). From above analysis, there are two stages for c ≥ 1 in the tensile process. When the strain is smaller than $\varepsilon_I$, the GNR-k can readily elongate up without some remarkable stresses. Therefore, the above second mechanism (that is, the thermal conductivity increases with increasing system length) dominate the thermal conductivity for small strain (<20% for GNR-k3nm and <60% for GNR-k5nm). If the strain is higher than $\varepsilon_I$, the tensile stress increases sharply with increasing of tensile strain before the final fracture. Hence, the first mechanism (that is, the tensile deformation causes the interphonon scattering) dominate the decreasing thermal conductivity. Based on the above analysis, GNR-k exhibits ductility properties and the thermal conductivity is robustness with tensile strain.

## Conclusions

In summary, the extreme reduction of the thermal conductivity by tailoring sizes in GNR-k was obtained using nonequilibrium MD simulations. The results show that the thermal conductivity of GNR-k (around 5.1 Wm$^{-1}$K$^{-1}$) is about two orders of magnitude lower than that of the corresponding graphene nanoribbon (GNR)(around 151.6 Wm$^{-1}$K$^{-1}$). The micro heat flux on each carbon atoms of GNR-k was provided in detail to clarify the reduction of its thermal conductivity, in which the elongation of real heat flux path, the overestimation of the real heat flux area and phonon scattering at the edge of vacancy have been identified as the three main reasons of the reduction. Moreover, the strain engineering effect on thermal conductivity of GNR-k and its thermal robust property was obtained in detail. Our results provide physical insights into the origins of the ultralow and robust thermal conductivity of GNR-k, which also suggests that the GNR-k can be used as a potential thermoelectric material and should be useful for nanoscale heat management.

\* Corresponding author.
*Email address*: (JZ) junhua.zhao@163.com; (JCZ) jczheng@xmu.edu.cn

## Appendix

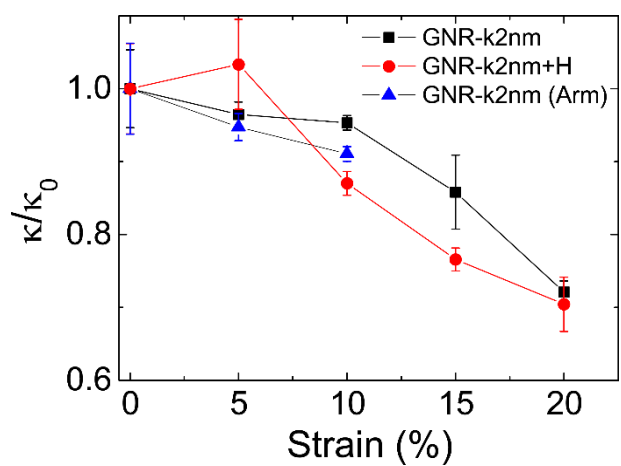

**Fig. A1** Thermal conductivity of graphene kirigami with/without Hydrogen atom terminated (solid circle in red and solid square in black), and a case of an armchair GNR-k (solid triangle in blue) with the same size are shown for a comparison.

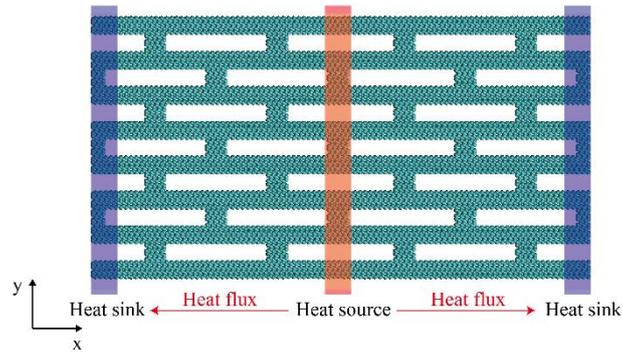

**Fig. A2** The schematic of GNT-k with vacancy shape parallel to the transport direction.

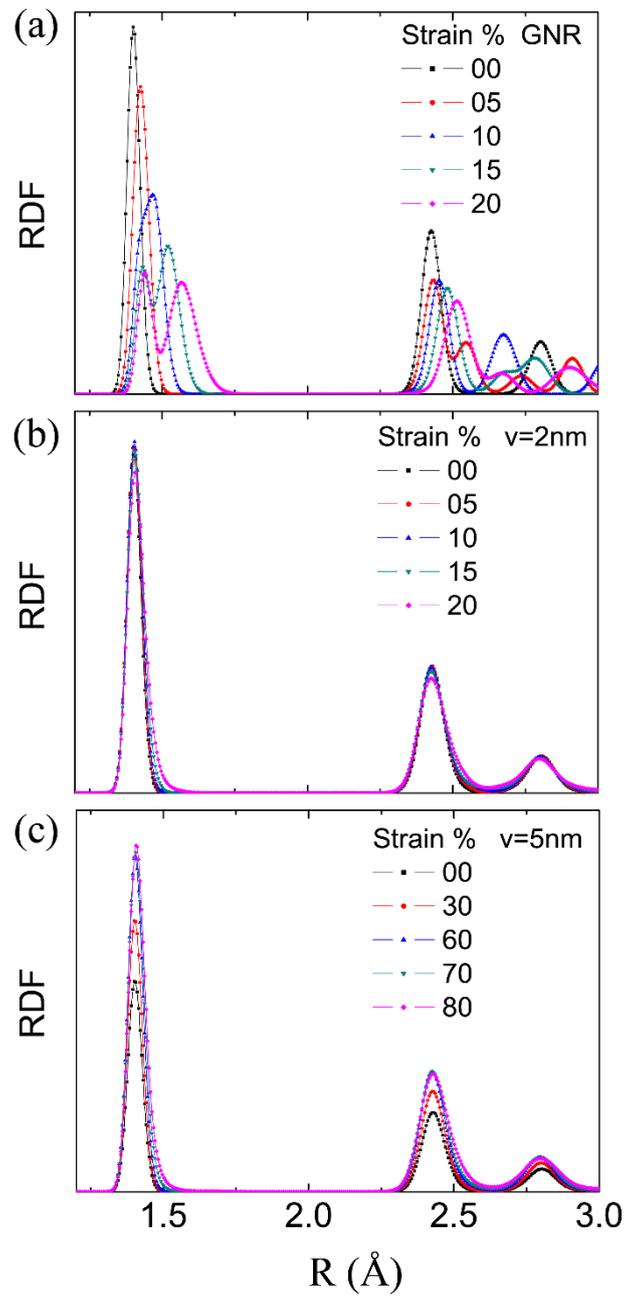

**Fig. A3** Radial distribution function of (a) GNR and (b) GNR-k2nm and (c) GNR-k5nm under different strain conditions.

# Thermal Conductivity of Graphene Kirigami with different Geometry Shapes and Strains

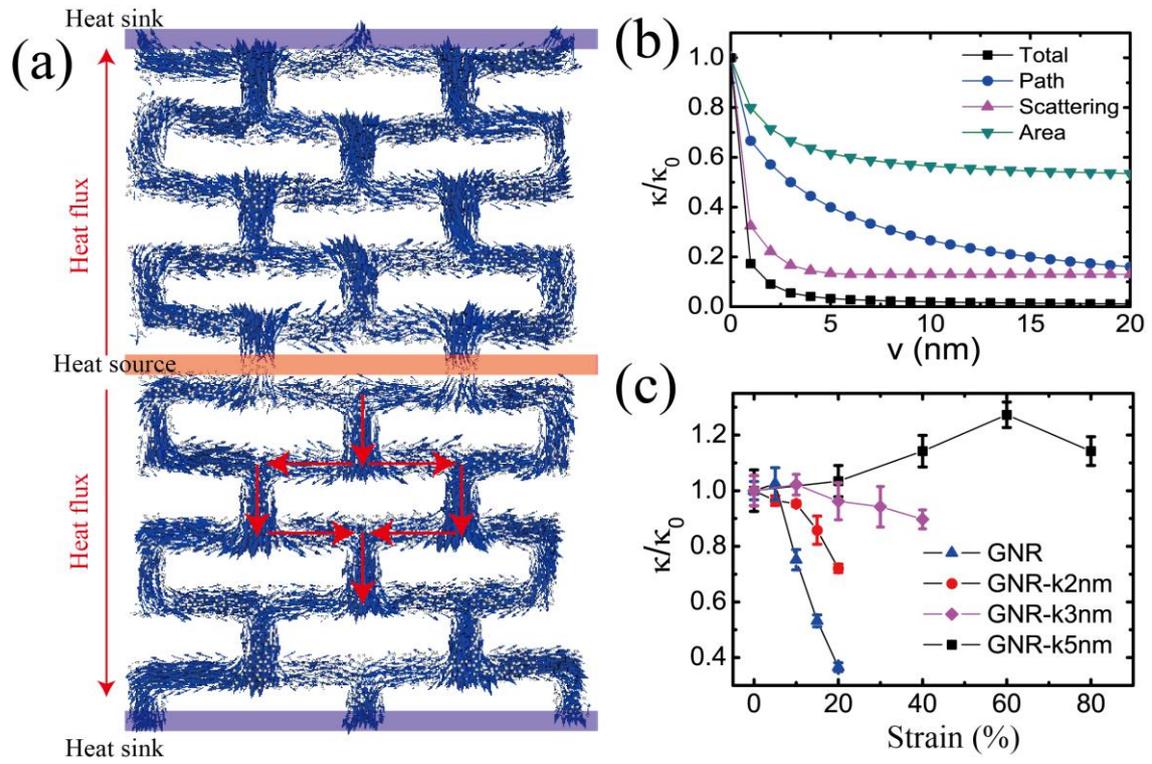

**Table of Contents Graphics**